\documentclass[10pt,journal]{IEEEtran}
\usepackage{url}            
\usepackage{booktabs}       
\usepackage{amsfonts}       
\usepackage{nicefrac}       
\usepackage{microtype}      
\usepackage{lipsum}
\usepackage{fancyhdr}       
\usepackage{graphicx}       

\usepackage{algorithm}
\usepackage{amsmath,amssymb,amsfonts}
\usepackage{algorithmic}
\usepackage{textcomp}
\usepackage{xcolor}
\usepackage{blindtext}
\usepackage{subfig}
\usepackage{cuted}
\usepackage{physics}
\usepackage{xcolor}
\usepackage[long]{optidef}
\usepackage[detect-none]{siunitx}
\usepackage{multirow}
\usepackage{array}

\ifCLASSOPTIONcompsoc
  \usepackage[nocompress]{cite}
\else
  \usepackage{cite}
\fi
\ifCLASSINFOpdf
\else
\fi
\begin{document}
%
\title{Optimized Topology Control for IoT Networks using Graph-based Localization}
%
%
%
%

\author{Indrakshi Dey,~\IEEEmembership{Senior Member,~IEEE,}
        and~Nicola Marchetti,~\IEEEmembership{Senior Member,~IEEE}
\thanks{I.~Dey is with the Walton Institute for Information and Communications Systems Science, South East Technological University, Ireland. (Email: indrakshi.dey@waltoninstitute.ie)}
\thanks{N. Marchetti is with CONNECT, School of Engineering, Trinity College Dublin, Ireland. (E-mail: nicola.marchetti@tcd.ie)}
\thanks{The contribution from I. Dey and N. Marchetti is supported by HORIZON-MSCA-2022-SE-01-01 project COALESCE under Grant Number 101130739, HORIZON-EOSC-03-2020 project QCloud under Grant Number INFRAEOSC-03-2020 and ERDF project SAtComm under Grant Number EAPA\_0019/2022.}
}

%
%

\markboth{Submitted to IEEE Transactions on Network Science and Engineering 2024}%
{Shell \MakeLowercase{\textit{et al.}}: Bare Demo of IEEEtran.cls for Computer Society Journals}
%



\IEEEtitleabstractindextext{%
\begin{abstract}
The key research question we are addressing in this paper, is how local distance information can be integrated into the global structure determination, in the form of network graphs realization for IoT networks. IoT networks will be pervading every walk of life over the next few years with the aim of improving quality of life and enhancing surrounding living conditions, while balancing available resources, like energy and computational power. As we deal with massive number of heterogeneous devices contributing to each IoT network, it is of paramount importance that the IoT network topology can be designed and controlled in such a way that coverage and throughput can be maximized using a minimum number of devices, while tackling challenges like poor link quality and interference. We tackle the above-mentioned problem of topology design and control through our designed graph-realization concept. End-nodes and gateways are identified and placed within neighborhood sub-graphs and their own coordinate system, which are stitched together to form the global graph. The stitching is done in a way that transmit power and information rate are optimized while reducing error probability. 
\end{abstract}

}

\maketitle

\IEEEdisplaynontitleabstractindextext

%
\IEEEpeerreviewmaketitle

\vspace{-3mm}
\section{Introduction}\label{sec:introduction}

%
%
%
%
The Internet of Things (IoT) is rapidly transforming our world, ushering in an era of ubiquitous connectivity and intelligent automation. By enabling the interconnection of billions of devices equipped with sensors and actuators, IoT empowers the creation of a vast network of physical objects capable of collecting, processing, and exchanging real-time data \cite{1}. This technological revolution is disrupting various sectors, fostering the development of smart applications across diverse domains. From smart cities with optimized traffic management and energy grids \cite{2} to intelligent healthcare systems that enable remote patient monitoring and personalized medicine \cite{3}, the potential impact of IoT is nothing short of transformative.  In the realm of industrial automation, IoT facilitates predictive maintenance, real-time process optimization, and increased production efficiency \cite{4}.

\subsection{Motivation}

Despite the immense potential of IoT, ensuring efficient and reliable data transmission in large-scale networks presents significant challenges. Unlike traditional communication networks designed for human users with high bandwidth demands (e.g., video streaming), IoT networks operate under fundamentally different constraints. A defining characteristic of IoT networks is the prevalence of resource-constrained devices. These devices often have limited battery power, computational capabilities, and memory. This necessitates the development of lightweight communication protocols and energy-efficient algorithms that can function effectively with minimal resources \cite{5,6}. Network designers need to carefully consider the trade-offs between data fidelity, transmission range, and power consumption to optimize network performance within these constraints.

Many IoT deployments involve dynamic network topologies with mobile sensors and actuators, requiring adaptive configurations to maintain connectivity and data integrity as conditions change. Routing protocols must be flexible to handle these shifts without compromising data delivery \cite{7}. The diverse ecosystem of IoT devices, with varying communication protocols and capabilities, creates challenges for network management and interoperability, making standardized protocols and resource management essential for seamless communication \cite{8}. Additionally, IoT networks often operate with limited bandwidth, particularly in Low Power Wide Area Networks (LPWANs), requiring efficient use of bandwidth and minimized data packet sizes for reliable communication \cite{9}. Security and privacy concerns also arise, necessitating secure communication and protection of sensitive data \cite{8,9}.

Traditional network optimization approaches have often focused on maximizing throughput or minimizing energy consumption \cite{10,11}. While these objectives are important for network performance, they might not be suitable for all IoT applications. For instance, in domains like industrial automation or remote healthcare monitoring, ensuring data fidelity and minimizing error probability become paramount.  Furthermore, with the ever-increasing scale and complexity of IoT networks, scalability becomes a critical design consideration. Algorithms and protocols need to be efficient and adaptable to handle large numbers of devices, while maintaining network performance and stability \cite{12}.

\vspace{-3mm}
\subsection{Related Works}

Database-matching (DB-M) has become a viable localization method in IoT networks by comparing real-time signal measurements (e.g., Received Signal Strength Indicator (RSSI), Time-of-Arrival (ToA), or Angle-of-Arrival (AoA)) with a pre-established database of signal fingerprints from known locations \cite{13}. Traditional DB-M techniques use probabilistic models or distance-based metrics to find the closest match \cite{14,15}. To enhance accuracy, machine learning (ML) methods have been applied. Key ML techniques include Artificial Neural Networks (ANN), which capture complex non-linear relationships \cite{16,17}, Random Forests for robustness against noise and overfitting \cite{18,19}, and Deep Reinforcement Learning (DRL), particularly useful in dynamic environments that require database updates \cite{20,21}.

While DB-M offers advantages for localization, it presents challenges for IoT deployments. A key issue is the need for a comprehensive signal fingerprint database, which requires extensive data collection campaigns. Maintaining this database in dynamic environments, where signal characteristics fluctuate, adds significant overhead \cite{13}. Additionally, ML-based DB-M methods often require computationally intensive training and inference \cite{16,20}, making them unsuitable for resource-constrained IoT devices. Pre-trained models may struggle to adapt to network or environmental changes if the database isn't regularly updated, leading to degraded localization accuracy over time.

The authors in \cite{22} are the first to combine node localization with topology control for power optimization, using noisy distance measurements and eigenvector synchronization. The network topology is constructed via linear programming to maximize spatial usage and throughput, emphasizing accurate localization for power-efficient IoT design. In \cite{23}, localized algorithms allow nodes to make decisions based on neighborhood information, utilizing techniques like iterative power adjustment and node state switching for energy conservation. Decentralized solutions are highlighted for scalability in large networks. In \cite{24}, Graph Signal Processing (GSP) is used to treat network measurements as graph signals, enabling spectral analysis and filtering to enhance localization and topology inference.

Hybrid approaches combine the strengths of different localization methods to overcome their individual limitations. Kalman Filtering and Bayesian techniques are often used to fuse data from various sources for probabilistic location estimates. A hybrid GPS-Inertial Measurement Unit (IMU) system has been proposed for indoor localization \cite{25}, where GPS provides accurate outdoor data, while IMUs track movement indoors where GPS is unreliable. Wi-Fi signal "fingerprints" combined with IMU data enhance accuracy and address signal fluctuations \cite{26}. Bluetooth Low Energy (BLE) beacons and Ultra-wideband (UWB) offer precise indoor positioning, with techniques like RSSI estimation, interference cancellation, and sensor data aggregation further improving accuracy \cite{27}.

The techniques mentioned above do not fully address challenges in resource-efficient IoT network topology design. Many IoT applications involve mobile nodes or dynamic environments, but state-of-the-art methods often assume static networks. Adapting algorithms to handle node movement and environmental changes is crucial. In dense deployments, interference between devices can degrade signal quality, and while hybrid methods mitigate this, they don’t handle large-scale interference explicitly. Additionally, IoT networks include devices with diverse capabilities, and techniques focused on homogeneous nodes may not optimize topologies effectively. Applications like self-driving cars and healthcare monitoring require real-time, low-latency decisions. Thus, localization and topology optimization must balance accuracy with computational efficiency, especially given the resource constraints of many IoT devices.

\vspace{-3mm}
\subsection{Contributions}

The primary contribution of the paper is to propose a novel algorithm, named IoTNTop, specifically designed to optimize power allocation and network topology within IoT networks, with a primary focus on minimizing the overall error probability, while maintaining energy efficiency. IoTNTop leverages a unique optimization objective function that integrates error probability, transmission range, energy considerations, and channel characteristics. This comprehensive approach allows IoTNTop to strike a balance between achieving low error rates and maintaining energy efficiency, which is crucial for battery-powered devices in IoT deployments. This work contributes to the field of IoT network optimization specifically through :
\begin{itemize}
    \item \textbf{Novel Graph Modeling Technique} - Segmenting a large network graph into sub-graphs and stitching breaks down large IoT networks into smaller sub-graphs, for computational efficiency and easier analysis. The ability to align and stitch these sub-graphs back together while addressing fold-overs is essential for constructing the entire network topology. Error-Aware Embedding techniques involving eigenvectors, least-square solutions, semi-definite programming (SDP) and multi-dimensional scaling (MDS) are used in combination. We show how these techniques improve accuracy and robustness against measurement noise, when determining the positions of both end-nodes and gateways.
    \item \textbf{Error-Centric Topology Optimization} - The paper introduces a novel optimization problem that prioritizes minimizing error probability while maximizing data transmission code rate (bits coded per symbol), differing from traditional methods that focus on energy efficiency. It formally integrates information-theoretic concepts like encoding/decoding, error probability, and code rate, offering a solid theoretical framework for analyzing and optimizing error performance. Additionally, power and range optimization accounts for realistic constraints such as node transmit power, signal propagation, path loss, and transmission range.
    \item \textbf{IoTNTop Algorithm} - The algorithm iteratively designs network topologies by prioritizing high-SNR links to enhance error resilience. Numerical analysis shows that IoTNTop is computationally efficient and handles larger IoT networks better than alternative methods. This is particularly beneficial for large-scale IoT deployments, such as smart cities, environmental monitoring, and industrial automation, where managing many interconnected devices is critical. By reducing the iterations needed for optimal solutions, IoTNTop lowers the computational complexity of network setup and management, making it ideal for real-world large-scale IoT applications.
\end{itemize}

The rest of the paper is organized as follows. Section II introduces the graph embedding algorithm designed to capture the intricate relationships within an IoT network, including spatial distribution and temporal dynamics of information flow. Section III delves into the critical question of how to achieve feasible power assignments, ensuring reliable communications by minimizing error probability across the network. Section IV unveils our novel algorithm IoTNTop, that iteratively refines the topology for optimal power assignment while prioritizing error minimization. Numerical analysis and results demonstrate IoTNTop's superiority in maximizing resource utilization, transmission code rate, and overall error resilience.
\begin{figure}[t]
\begin{center}
    \includegraphics[width=0.99\columnwidth]{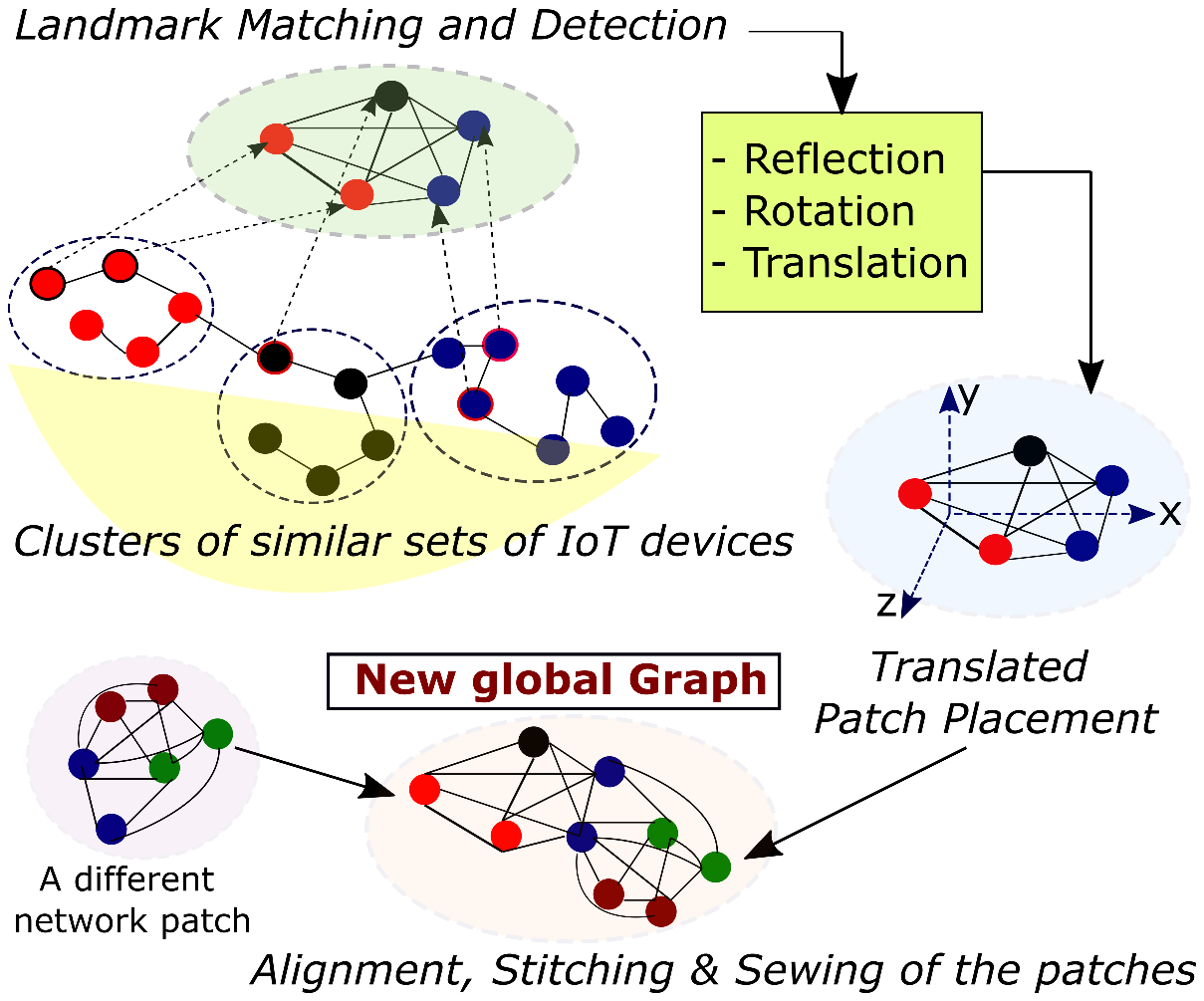}
\end{center}
\vspace{-52mm}
\caption{A conceptual overview of proposed optimized topology control. Different colors represent different types of IoT nodes.}
\vspace{-5mm}
\label{FIG2a}
\end{figure}

\section{Network Graph Modelling}

Let us consider an IoT network with $l$ end-nodes (consisting of sensors, trackers, meters, actuators etc.) within the set $\mathcal{L}$, and $s$ gateways (central nodes) within the set $\mathcal{S}$. In the first step, we solve the localization and embedding problem of the IoT end-nodes, i.e., we look into the set $\mathcal{L}$ first. Let a set of $|V| = l$ IoT end-nodes be represented as an oriented graph $G = (V,E)$ with $E$ being the set of the links between the IoT end-nodes forming the network, and $|E| = m$. Each link is associated with an Euclidean distance measurement $d_{aj}$ between nodes $a$ and $j$, $(a,j) \in E$ and $d_{aj} = d_{ja}$. The main goal of our work is to design an IoT network as a graph that maximizes area coverage with minimal nodes, while optimizing network throughput, link quality, and signal-to-noise ratio (SNR), and minimizing error probability. To achieve this, we first divide the network graph into smaller sub-graphs. Sub-graphs with more than $k$ shared nodes are reflected, rotated, translated, and aligned to remove fold-overs, which helps visualize patterns and trends that may be obscured in the full dataset \cite{28}.

Fig.~\ref{FIG2a} presents a visual representation of topology optimization for integrating disparate clusters of IoT nodes into a cohesive heterogeneous network. It involves four key steps.
\begin{itemize}
    \item First, we identify common reference points or features (landmarks) across heterogeneous clusters of IoT devices. These landmarks are based on device attributes, functionalities and communications patterns through specific transformations, including reflection, rotation and translation.
    \item After landmark matching, individual clusters are treated as ``patches" and are adjusted to align with each other based on the identified landmarks. This alignment process may involve translating (shifting) the patches in the $\mathsf{x}$, $\mathsf{y}$, or $\mathsf{z}$ dimensions to ensure proper overlap.
    \item Once the patches are aligned, they are combined or stitched together to create a unified representation. This step involves merging models associated with each cluster, taking into account any overlaps and redundancies.
    \item The final result of this process is a new global graph that integrates information from all the original clusters. This graph represents relationships between devices and communication patterns.
\end{itemize}

\subsection{Graph-embedding for End-nodes}

Let the sub-graphs formed after breaking the graph $G$ be denoted by $P_1, \dotso, P_N$, where $N$ is the total number of broken graphs. Let the set of the sub-graphs formed after alignment and stitching be given by $p_1, \dotso, p_n$, where $d_{aj} = ||p_a - p_j|| + \varepsilon_{aj}$ where $\varepsilon_{aj}$ is the added noise due to breaking and stitching back. It is worth-mentioning here that $P_a$ consists of all the nodes within 1-hop neighborhood of the $a$th IoT end-node. To analyze interactions between two nodes, we can find the intersection of their 1-hop neighborhoods. This intersection forms the vertex set of the subgraph $G(a,j)$, where $(a,j) \in E(G(a,j))$ is an edge in the original graph. So considering the nodes $a$ and $j$, let us assume that we want to align the sub-graphs $P_a$ and $P_j$. In order to achieve that, the first step is to formulate the relative reflection, $z_{aj} \in \{-1,+1\}$.

\subsubsection{Sub-graph Reflection}

We first build a $N \times N$ sparse symmetric matrix $\mathcal{Z}$, where,
\begin{align}
    \mathcal{Z} = (z_{aj}) =
  \begin{cases}
    1   & \text{No reflection needed for alignment}\\
    -1  & \text{One set of reflection needed}\\
    0  & \text{Cannot be aligned}
  \end{cases}
\end{align}
We compute the top eigenvector $v^{\zeta}_1$ of $\zeta = \Delta^{-1}\mathcal{Z}$ with $\Delta$ as the diagonal matrix $\Delta_{\alpha \alpha} = \text{deg}(\alpha)$, $\zeta v^{\zeta}_1 = \lambda^{\zeta}_1 v^{\zeta}_1$ and estimate,
\begin{align}
\hat{z}_a = \text{sign}(v^{\zeta}_1(a)) = v^{\zeta}_1(a)/|v^{\zeta}_1(a)|
\end{align}
and if $\hat{z}_a = - 1$, $P_a$ needs to be reflected to its mirrored version. The next step is to calculate the rotation of $\theta_{aj}$ between $P_a$ and $P_j$. 

\subsubsection*{Construction of $\mathcal{Z}$}

We can start with an $N \times N$ empty matrix. For each pair of subgraphs $P_a$ and $P_j$, we calculate the correlation between the available distance measurements associated with the shared nodes in $P_a$ and $P_j$, using the concept of Pearson correlation coefficient, as,
\begin{align}
    \tilde{\mathsf{r}} = \frac{\sum_{i = 1}^n (d_{ai} - \mu_a)(d_{ji} - \mu_j)}{(n - 1)\alpha_a \alpha_j}
\end{align}
where $\tilde{\mathsf{r}}$ is the Pearson correlation coefficient between the distance measurements associated with the shared nodes in subgraphs $P_a$ and $P_j$, quantifying the strength and direction of the linear relationship between these two sets of distances, $d_{ai}$ is the Euclidean distance measurement between node '$a$' in subgraph $P_a$ and the $i$th shared node, $\mu_a$ is the mean of the distance measurements $d_{ai}$ for all shared nodes '$i$' in subgraph $P_a$, $d_{ji}$ is the Euclidean distance measurement between node '$j$' in subgraph $P_j$ and the $i$th shared node, $\mu_j$ is the mean of the distance measurements $d_{ji}$ for all shared nodes '$i$' in subgraph $P_j$, $n$ is the number of shared nodes between subgraphs $P_a$ and $P_j$, $\alpha_a$ is the standard deviation of the distance measurements $d_{ai}$ in subgraph $P_a$, and $\alpha_j$ is the standard deviation of the distance measurements $d_{ji}$ in subgraph $P_j$.

The equation calculates the correlation coefficient by first centering the distance measurements for each subgraph by subtracting their respective means. Then, it computes the product of these centered distances for each shared node and sums these products across all shared nodes. Finally, it normalizes this sum by the product of the standard deviations and $n - 1$ to obtain the correlation coefficient. The correlation coefficient helps determine if two subgraphs need to be reflected for alignment during the graph embedding process. If the correlation is sufficiently negative (below a threshold, typically -0.5), it indicates a need for reflection. If $\tilde{\mathsf{r}} < - 0.5$, $z_{aj} = -1$, else $z_{aj} = 1$, The reliability of the method depends on the chosen value of -0.5 which has been chosen judiciously through trial and error, and the expected accuracy of distance measurements observed in practical IoT networks.

\subsubsection{Sub-graph Rotation}

We construct the $N \times N$ Hermitian matrix $R = (r_{aj})$ such that,
\begin{align}
 r_{aj} =
  \begin{cases}
    e^{\imath \theta_{aj}} & \quad P_a~\text{\&}~P_j~\text{can be aligned}\\
    0  & \quad P_a~\text{\&}~P_j~\text{cannot be aligned}
  \end{cases}
\end{align}
where $\theta_{aj} \in [0,2\pi)$, $\theta_{aj} = - \theta_{ja} \mod 2\pi$, $r_{aj} = \bar{r}_{ja}$, for any complex number and its complex conjugate. We compute the top eigenvector $v_1^{\mathcal{R}}$ of $\mathcal{R} = \Delta^{-1}R$ with $\mathcal{R} v^{\mathcal{R}}_1 = \lambda^{\mathcal{R}}_1 v^{\mathcal{R}}_1$ and estimate,
\begin{align}
e^{\imath \hat{\theta}_{a}} = v^{\mathcal{R}}_1(a)/|v^{\mathcal{R}}_1(a)|
\end{align}
to rotate the patch $P_a$. The next step is to compute the translation of the sub-graph $P_b$ into aligned sub-graph $P_a$. Here $P_b$ is the $b$th patch or subgraph.

\subsubsection*{Construction of $R$}

We start with an $N \times N$ matrix where initially all elements are $0$.  For each pair of subgraphs $P_a$ and $P_j$, where alignment is possible, we calculate $\theta_{aj}$ following the Landmark Matching procedure \cite{29}. Let $\bar{X} = \{\bar{x}_1, \bar{x}_2, \dotso, \bar{x}_m\}$ with $\mu_{\bar{X}} = \frac{1}{m}\sum_{\bar{i} = 1}^m \bar{x}_{\bar{i}}$ as the mean points in $\bar{X}$, represent the coordinates of $m$ shared landmarks in $P_a$, $\bar{Y} = \{\bar{y}_1, \bar{y}_2, \dotso, \bar{y}_m\}$, represent the corresponding coordinates in $P_j$ with $\mu_{\bar{Y}} = \frac{1}{m}\sum_{\bar{i} = 1}^m \bar{Y}_{\bar{i}}$ as the mean points in $\bar{Y}$, $\bar{z}_i = \bar{x}_i - \mu_{\bar{X}}$ is the zero-centered coordinate for landmark $\bar{i}$ in $P_a$, $\bar{w}_i = \bar{Y}_i - \mu_{\bar{Y}}$ is the zero-centered coordinate for landmark $\bar{i}$ in $P_j$, and $\mathsf{H} = \sum_{\bar{i} = 1}^m \bar{z}_i \bar{w}_i$ is the covariance matrix between $\bar{X}$ and $\bar{Y}$.  

Now let us define,
\begin{align}
    \text{Rotation :}~&\bar{y}_i \approx e^{\imath\theta} \bar{x}_i = (\cos\theta + \imath\sin\theta) \bar{x}_i\\
    \text{Translation :}~&\bar{y}_i \approx e^{\imath\theta} \bar{x}_i + \bar{t}\\ 
    \text{Scaling :}~&\bar{y}_i \approx \bar{\mathsf{s}}e^{\imath\theta} \bar{x}_i + \bar{t} 
\end{align}
where $\bar{t}$ is the translation vector and $\bar{\mathsf{s}}$ is the scaling factor. The goal is to find the parameter values that minimize the discrepancy between the transformed landmarks from $P_a$ and their corresponding points in $P_j$. We formulate this as a least-squares problem; 
\begin{align}\label{eq4a}
    \text{Minimize}_{\mathsf{s}, \bar{t}} :~\sum_{\bar{i} = 1}^m ||\bar{\mathsf{s}}e^{\imath\theta} \bar{x}_i + \bar{t} - \bar{y}_i||^2
\end{align}

In order to remove the effect of global translation, we substitute $\bar{z}_i$ and $\bar{w}_i$ for the centered coordinates in (\ref{eq4a}) to obtain the optimization problem as,
\begin{align}\label{eq4c}
    \text{Minimize}_{\mathsf{s}, \bar{t}} :~\sum_{\bar{i} = 1}^m ||\bar{\mathsf{s}}e^{\imath\theta} (\bar{x}_i - \mu_{\bar{X}}) + (\bar{t} - \mu_{\bar{Y}}) - (\bar{y}_i - \mu_{\bar{Y}})||^2
\end{align}
We solve the optimization problem using singular value decomposition (SVD) for the covariance matrix $\mathsf{H}$, to extract the value of $\theta$ using trigonometric functions and identities and then solve for $e^{\imath\theta}$ as a function of eigenvectors and eigenvalues of $\mathsf{H}\mathsf{H}^T$. Optimality of (\ref{eq4c}) is detailed in Appendix A.

\subsubsection{Sub-graph Translation}

For $P_b$, we compute,
\begin{align}\label{eq5}
p_a = p_a^{(b)} + t^{(b)}; i \in V_a; b = 1, \dotso, N
\end{align}
where $t^{(b)}$ is the associated translation. Using the over-determined system of (\ref{eq5}), we can estimated the global coordinates, $p_1, \dotso, p_n$ using least square solution to (\ref{eq5}). The set of translations $t^{(1)}, \dotso, t^{(N)}$ is disregarded. So for each edge $(a,b) \in E_b$, $p_a - p_j = p_a^{(b)} - p_j^{(b)}$. If we replace the patches with the $(\mathsf{x},\mathsf{y})$ coordinates, $\mathsf{x}_a - \mathsf{x}_j = \mathsf{x}_a^{(b)} - \mathsf{x}_j^{(b)}; \mathsf{y}_a - \mathsf{y}_j = \mathsf{y}_a^{(b)} - \mathsf{y}_j^{(b)}$, which can be solved separately. 

Now we can write, $\tau \mathsf{x} = \gamma^{\mathsf{x}}$, where $\tau$ is the least square translation matrix. It is constructed based on the relationships between the local coordinates of nodes in different subgraphs and their corresponding global coordinates. $\mathsf{x}$ is the $n \times 1$ vector of $\mathsf{x}$-coordinates of all IoT end-nodes localized in a sub-graph. The $n \times 1$ vector represents the $\mathsf{x}$-coordinates of all the IoT end-nodes that have been localized within a particular subgraph. The goal is to estimate these global $\mathsf{x}$-coordinates. $\gamma^{\mathsf{x}}$ is the vector with entries from the right-hand side of (\ref{eq5}). The vector containing the right-hand side values of the linear equations is derived from the subgraph translation process. These values are based on the local coordinates and the estimated translations between subgraphs. Similarly, we can write $\tau \mathsf{y} = \gamma^{\mathsf{y}}$. The accuracy of these estimated coordinates is then evaluated by comparing them to the true coordinates using an error metric. Now by adding all the equations corresponding to the same edge $(a,j)$ from different patches,
\begin{align}\label{eq6}
&\sum_{k \in \{1, \dotso, N\}; (a,j) \in E_b} \mathsf{x}_a - \mathsf{x}_j \nonumber\\
&= \sum_{k \in \{1, \dotso, N\}; (a,j) \in E_b} \mathsf{x}_a^{(b)} - \mathsf{x}_j^{(b)},~(a,j) \in E 
\end{align}
and similarly for the $\mathsf{y}$-coordinates we can formulate the least-square translation matrix $\tau$, i.e. $m \times n$ over-determined system of linear equations. If the estimated transition matrix is denoted by $\hat{\tau}$, the least square solutions to $\hat{\tau}\mathsf{x} = \gamma^{\mathsf{x}}$ and $\hat{\tau}\mathsf{y} = \gamma^{\mathsf{y}}$ will be given by, $\hat{p}_1, \dotso, \hat{p}_n$. The error in any estimated sub-graph coordinates can be calculated as $||p_a - \hat{p}_a||$. Now that we have solved the localization problem for IoT end-nodes, we can find the union of set $\mathcal{L}$ (end-nodes) and set $\mathcal{S}$ (gateways) to determine the locations of all nodes within the network.

\subsection{Graph-embedding for Central and End-nodes}

Let us assume that we want to connect the $i$th gateway, $i = 1, \dotso, s$ with the $j$th end-node, $j = 1, \dotso, l$. Therefore the Euclidean distance is $d_{ij}$ between the $i$th gateway and the $j$th end-node, where $(i,j) \in \mathcal{S} \cup \mathcal{L}$ and $\bar{E}(\mathcal{L}, \mathcal{L})$ are the set of edges between end-node and gateway, and two end-nodes, respectively. We now form a new network graph, $\bar{G} = (\bar{V},\bar{E})$ with $\bar{V} = \mathcal{S} \cup \mathcal{L}$ of size $|\bar{V}| = l + s$ and edge set $|\bar{E}| = m + \bar{m}$ where $\bar{m}$ is the number of edges between the end-nodes and gateways of the network. In this case, the partial distance measurement matrix is given by, $\mathbf{D} = \{d_{ij}:(i,j)\in \bar{E}(\mathcal{L}, \mathcal{L}) \cup \bar{E}(\mathcal{L}, \mathcal{S})\}$ and the graph realization problem becomes,
\begin{align}\label{eq7}
||\mathsf{x}_i - \mathsf{x}_j||^2_2 &= d^2_{ij}~\text{for}~(i,j) \in \bar{E}(\mathcal{L},\mathcal{S}) \nonumber\\
||\mathsf{x}_j - \mathsf{x}_a||^2_2 &= d^2_{ja}~\text{for}~(j,a) \in \bar{E}(\mathcal{S},\mathcal{S}) \nonumber\\
&\text{for}~i = 1, \dotso, s, j = 1, \dotso, l, a = 1, \dotso, l.
\end{align}

We start by randomizing the realization of the sub-graph embeddings $p_1, \dotso, p_n$. We also introduce a small positive constant of $\varrho$ (for example, $\varrho = 10^{-5}$). Once we have the end-node network mapping in place, we look at the network between the end-nodes and gateways by solving the semi-definite programming (SDP) problem,
\begin{align}\label{eq8}
\text{Maximize}~0,~\to~\text{subject to},~(\mathbf{0}; e_i-e_j)(\mathbf{0};e_i-e_j)^T \chi = d^2_{ij}
\end{align}
for $(i,j) \in \bar{E}(\mathcal{L},\mathcal{S})$ and $\mu \in \kappa^{l + s}$ where $e_i$ is an all-zeros vector with $i$th entry of 1 and 
\begin{align}\label{eq9}
\kappa^{l + s} = \Bigg\{\chi_{(l+s)\times (l+s)}~\Bigg|~\chi = \begin{bmatrix}
I_s & \mathsf{X}\\
\mathsf{X}^T & \mathsf{Y}
\end{bmatrix} \geq 0\Bigg\}
\end{align}
where $\chi$ is the matrix representation of $(\mathsf{X},\mathsf{Y})$ coordinates. If the vector $\omega$ represents the diagonal elements of the matrix $\mathsf{Y} - \mathsf{X}\mathsf{X}^T$, we compute the new subset of vertices $\tilde{V} \in \bar{V}\backslash \mathcal{S}$ such that $\omega_i < \varrho$ and form the end-node -gateway sub-graph, $\tilde{G} = (\tilde{V}, \tilde{E})$. We formulate the solution to (\ref{eq8}) as SDP matrices, as eigenvalues being non-negative means the resulting computed `distances' will make physical sense and SDP matrices generally enforce the triangle inequality, which is essential for realistic distances. In (\ref{eq8}), $\text{Maximize}~0$ indicates that the optimization problem aims to maximize the objective function while ensuring that all constraints are satisfied. The '$0$' in this context is a placeholder or represent a null objective function, implying that the primary focus is on feasibility rather than optimizing a specific numerical value. The actual optimization occurs through the constraints, which enforce conditions on the variables involved. Optimality of (\ref{eq8}) is detailed in Appendix~B.

\subsubsection*{Insights into $\chi$} 
$\chi$ is a block matrix representation of the coordinates of both end-nodes and gateways. It has dimensions $(l + s) \times (l + s)$. Here $I_s$ corresponds to fixed gateway node positions and is an $s\times s$ identity matrix, $\mathsf{X}$ is an $s \times l$ matrix where each column will hold the coordinates of an end-node to be estimated, $\mathsf{X}^T$ is the transpose of the end-node coordinate matrix, and $\mathsf{Y}$ is the $l \times l$ matrix where the diagonal elements will hold the squared pairwise distances between the end-nodes while the off-diagonal elements will be filled during the solution process. 

Let us consider a small network with 2 gateways and 3 end-nodes. Here $\chi$ will look like,
\begin{align}\label{eq9a}
\chi = \begin{bmatrix}
1 & 0 & \mathsf{x}_1 & \mathsf{x}_2 & \mathsf{x}_3\\
0 & 1 & \mathsf{y}_1 & \mathsf{y}_2 & \mathsf{y}_3\\
\mathsf{x}_1 & \mathsf{y}_1 & \cdot & \cdot & \cdot\\
\mathsf{x}_2 & \mathsf{y}_2 & \cdot & \cdot & \cdot\\
\mathsf{x}_3 & \mathsf{y}_3 & \cdot & \cdot & \cdot
\end{bmatrix} 
\end{align}

The SDP formulation, together with specific structure of $\chi$, allows to optimize the coordinates of the end-nodes to best match with the available set of distance measurements. Crucially, our formulated method assumes that we don't have direct pairwise distance measurements for all possible node pairs. Also, we assume that our network is connected, or at least has connected components that are large enough to be embedded. Finally, the SDP formulation is more robust to noisy measurements by adding regularization terms or relaxing the equality constraints in (\ref{eq8}).

\subsection{Sub-graph Embedding for the Global Graph}

For sub-graph embedding in our globally non-rigid IoT network graph, we resort to the three-step method outlined in \cite{22}. \footnote{It is worth-mentioning here, that the term 'rigid' refers to the property of a graph or network where the relative distances between all nodes remain constant. In other words, a rigid graph maintains its shape and structure even when subjected to transformations like rotations or translations. In this work, we resort to a non-rigid approach for sub-graph embedding because we are looking at topology optimization of IoT networks in a dynamic scenario.} The first step is to estimate the distance, $d'_{ij}$ with $(i,j) \notin E_k$,\footnote{We are considering an annular graph, with a network where the IoT devices are distributed randomly on an annulus surrounding the IoT gateway at the center.} $d'_{ij} = (\underline{d_{ij}} + \overline{d_{ij}})/2 $,
for,
\begin{align}\label{eq1}
\overline{d_{ij}} &= (\overline{d_{aj}} + \overline{d_{ai}})/2; \nonumber\\
\overline{d_{aj}} &= \min_{k:(a,k),(j,k) \in E_k} d_{ak} + d_{jk};~\overline{d_{ai}} = \min_{k:(a,k),(i,k) \in E_k} d_{ak} + d_{ik}\\
\underline{d_{ij}} &= \max\bigg[\max_{k:(a,k)\in E_k}d_{ak}, \max_{k:(i,k)\in E_k}d_{ik}, \max_{k:(j,k)\in E_k}d_{jk}\bigg]
\end{align}

The second step is to compute local coordinates of all nodes in a sub-graph using multi-dimensional scaling \cite{24} on the complete set of pairwise distances. We can express, $\mathbf{D} = -1/2 \mathbf{J}\mathbf{L}\mathbf{J}$ where $\mathbf{J} = \mathbf{I}_{s+l} - 1/(s+l) \mathbf{1}\mathbf{1}^t$, $\mathbf{1}$ is a matrix of ones, $\mathbf{I}$ is the identity matrix, $\mathbf{L}$ is the matrix of squared pairwise distances, $\mathbf{L} \in \mathbb{R}^{(s+l) \times (s+l)}$, $t$ denotes the transpose, $s+l \geq 1$ is an integer, and the matrix of the local coordinates of all nodes in a patch can be calculated as,
\begin{align}\label{eq2}
P_k =~\mathbf{U}_k \sqrt{\Lambda_k} &= \bigg[\sqrt{\lambda_1}\mathbf{u}_1, \dotso, \underbrace{\sqrt{\lambda_{f+1}}\mathbf{u}_{f+1}}_{= 0}, \dotso, \underbrace{\sqrt{\lambda_k}\mathbf{u}_k}_{= 0}\bigg] \nonumber\\
&\in \mathbb{R}^{(s+l) \times k}
\end{align}
where $({\lambda_{(s+l)}},\mathbf{u}_{(s+l)})$ are the eigen-pairs of the $\mathbf{D}$ matrix.

The final step is to refine the embedding using iterative majorization technique. The coordinates of each IoT node are updated according to,
\begin{align}\label{eq3}
p_i \leftarrow 1/~\text{deg}_i(P_k) &\sum_{j \in V_k, (a,i,j) \in E_k} [p_j + d_{ij}(p_j - p_a)\nonumber\\
&\times \text{inv}~(||p_j - p_a||)]
\end{align}
where $\text{deg}_i(P_k)$ denotes the degree of node $i$ in patch $P_k$ and,
\[ \text{inv}~(\mathsf{x}) =
  \begin{cases}
    1/\mathsf{x}       & \quad \text{if } \mathsf{x} \neq 0\\
    0  & \quad \text{if } \mathsf{x} = 0
  \end{cases}
\]

\subsection{Aligning Sub-graphs and Sewing them together}

Now after finding the embedding for the sub-graph $P_k$, we need to combine under the same objective function, the contribution of the end-node-to-end-node communications and the end-node-to-gateway communications. For the above, we formulate a least-square-problem-based synchronization as,
\begin{align}\label{eq4}
    \min_{\mathsf{x}} \sum_{(i,j)\in \bar{E}} (\mathsf{x}_i - \chi_{ij}\mathsf{x}_j)^2 = \min_{\mathsf{x}}~{\mathsf{x}}^T (\mathbf{D} - \chi)\mathsf{x}
\end{align}
where,
\begin{align}
\chi = \begin{bmatrix}
\mathcal{\tilde{L}} & \mathcal{\tilde{S}}\\
\mathcal{\tilde{S}}^T & \mathcal{W}
\end{bmatrix},~
\mathbf{D} = \begin{bmatrix}
\mathbf{D}_{\mathcal{\tilde{L}}} & 0\\
0 & \mathbf{D}_{\mathcal{W}}
\end{bmatrix}
\end{align}
where $\mathbf{D}_{\mathcal{\tilde{L}}}$ is the pairwise distance matrix within the set $\mathcal{L}$, $\mathbf{D}_{\mathcal{W}}$ is the pairwise distance matrix within the set $\mathcal{W}$. The sets, $\mathcal{\tilde{L}}_{c \times c}$, $\mathcal{\tilde{S}}_{c \times k}$ and $\mathcal{W}_{k \times k}$ denote the end-node-to-end-node, end-node-to-gateway and gateway-to-gateway measurements. We will use the Lagrangian multiplier $\mu$ concept to solve the minimization problem of (\ref{eq4}) as,
\begin{align}\label{eq11}
    \chi^* = \big(\mathbf{D}_{\mathcal{\tilde{L}}} - \mathcal{\tilde{L}} + \mu \mathbf{I}\big)^{-1}(\tilde{\mathcal{S}}\sigma)
\end{align}
where $\sigma$ denotes the sign of the gateway vertices, $+$-ve sign represents similar or cooperative vertices\footnote{gateway vertices that have some degree of functional redundancy or complementary roles within the network}, $-$-ve sign represents dissimilar or competitive vertices\footnote{gateway vertices that have conflicting roles or compete for resources within the network}. In Algorithm 1, we summarize our proposed algorithm for graph embedding within a massive IoT network with multiple heterogeneous IoT nodes and more than one central node or gateway.

\begin{algorithm}[h]
\caption{~~Graph Embedding within a massive IoT Network}\label{tab:x}
\textbf{Input} : Graph $G=(V,E)$ representing the IoT network (end-nodes and gateways)\\
Distance measurements $d_{ij}$ between connected nodes $(i,j)$\\
\textbf{Step 1 - Break Down into Sub-graphs} : Divide the network graph $G$ into smaller sub-graphs $P_1, P_2, \dotso, P_N$.\\
Ensure that sub-graphs with more than $k$ nodes in common are considered for later alignment.\\
\textbf{Step 2 - Graph Embedding for End-nodes} : For each sub-graph $P_a$\\
\noindent\textbf{Sub-graph Reflection:}\\
Build sparse symmetric matrix $\mathcal{Z}$. Entries are determined using Pearson correlation on available distance measurements:
\begin{itemize}
    \item $z_{aj} = 1$ (no reflection) if $\tilde{\mathsf{r}} \geq 0.5$
    \item $z_{aj} = -1$ (reflection needed) if $\tilde{\mathsf{r}} < -0.5$
    \item $z_{aj} = 0$ (cannot align) otherwise
\end{itemize}
Compute the top eigenvector $v^{\zeta}_1$ of $\zeta = \Delta^{-1}\mathcal{Z}$ \\
Estimate $\hat{z}_a =$ sign$(v^{\zeta}_1(a))$. 
If $\hat{z}_a = -1$, reflect the sub-graph.
Build Hermitian matrix $R$ using Landmark Matching.
Compute top eigenvector $v_1^{\mathcal{R}}$ of $\mathcal{R}$.
Estimate $e^{\imath\hat{\theta}_a} = v_1^{\mathcal{R}}(a)/|v_1^{\mathcal{R}}(a)|$ to determine rotation angle.\\
\noindent\textbf{Sub-graph Translation:}\\
Using least-squares on over-determined system of linear equations, compute global coordinates $p_1, \dotso, p_n$.\\
\textbf{Step 3 - Landmark Matching} : For each pair of alignable sub-graphs $P_a, P_j$ :
Identify shared landmarks. Compute covariance matrix $\mathsf{H}$. Use singular value decomposition (SVD) on $\mathsf{H}$ to extract rotation angle $\theta$.\\
\textbf{Step 4 - Graph Embedding for End-nodes and Gateways} : \\
Form a new graph $\bar{G} = (\bar{V},\bar{E})$, including both end-nodes and gateways. Solve a SDP problem to find the embedding while respecting measured distances between end-nodes and gateways. \\
\textbf{Step 5 - Global Graph Embedding} : \\
For pairs of nodes without direct measurements; estimate distance $d'_{ij}$ using shortest paths within existing sub-graphs. For each sub-graph: Apply MDS to compute local coordinates using estimated pairwise distances. Refine the embedding: Use iterative majorization for refinement. \\
\textbf{Step 6 - Aligning and Sewing Together Sub-graphs} : 
Formulate the alignment as a least-squares synchronization problem, taking into account node-to-node and node-to-gateway distances. Solve the optimization to get final coordinates for all nodes.\\
\textbf{Output} : Coordinates $(\mathsf{x}_i, \mathsf{y}_i)$ for each node $i$ in the network
\end{algorithm}

Let us consider the true distance measurement between the $j$th and $a$th subgraphs as $\delta_{aj} = ||p_j - p_a||$ with random noise of $\epsilon_{aj}$ added to the distance measurement distributed uniformly over the range $[-\eta\delta_{aj}, \eta\delta_{aj}]$. Consequently, we can express the noisy distance measurements as $d_{aj} = \delta_{aj} + \epsilon_{aj}$. We consider networks of size 20, 50 and 100 nodes with average node degree of 14 - 20 and noise levels up to 70\% ($\eta = 0.7$ corresponding to 70\%).

We consider a network with gateways within a range of $\tilde{\rho}$ of different subgraphs. Let the true coordinates of all the nodes within subgraphs be represented by a $2 \times n$ matrix, $P = (p_1, \dotso, p_n)$ with estimated coordinates of $\hat{P} = (\hat{p}_1, \dotso, \hat{p}_n)$. We can compute the error in global localization of the nodes and the gateway owing to noisy measurement as, 
\begin{align}\label{eq18}
\mathcal{A}_{e} = \frac{\sqrt{\sum_{j = 1}^n ||p_j - \hat{p}_j||^2}}{\sqrt{\sum_{j = 1}^n ||p_j - p_0||^2}} = \frac{||P - \hat{P}||_F}{||P - p_0\mathbf{1}^t||_F}
\end{align}
where $||\cdot||_F$ denotes the Frobenius norm, $t$ denotes transpose, $\mathbf{1}$ denotes matrix containing all `1' elements, and $p_0 = \frac{1}{n}\sum_{j =1}^n p_j$ represents the center of mass of all the true coordinates of the subgraphs. The denominator in (\ref{eq18}) is the scaling and normalization factor for the graph embedding from the network perspective, representing the `spread' of the true node coordinates around their center of mass. This makes the error metric less sensitive to mere scaling differences. Both the numerator and denominator include a subtraction of the center of mass ($p_0$). This makes the metric largely insensitive to uniform translations of the entire embedding. By factoring out scale and translation, (\ref{eq18}) concentrates on measuring the error in the relative placement of nodes, which is often the key goal in graph embedding tasks. 

The primary goal of graph embedding is to capture the relationships between nodes based on their relative distances and positions, with less emphasis on the absolute location of the entire graph. By subtracting the center of mass, the focus shifts to errors in relative node placement rather than overall graph positioning. Shifting the entire set of estimated coordinates ($\hat{P}$) should not significantly affect the error metric ($\mathcal{A}_{e}$). Aligning both the true ($P$) and estimated ($\hat{P}$) coordinates with their respective centers ensures that the metric emphasizes relative positioning over absolute location.



\section{Topology Design and Control}

Next we formulate the topology extraction for an IoT network as an optimization problem with the target of minimizing error probability and maximizing information transmission rate over the link between the IoT nodes and the network gateways. In the proposed mathematical model, we also consider i) the transmit power from the node's side and received power at the gateway end, ii) how much energy is consumed when an IoT node fires its signal with a certain set of symbol coding on the information (the symbol coding can be simple space-time coding or other more complicated coding techniques like turbo coding) and iii) how much energy is consumed on the gateway side for absorbing the received signal. We also consider the link quality as a defining factor using the pathloss exponent, the transmission range of the node, and the far-field reception range for the gateway. However, we do not consider the energy consumption for accumulating and computing information at the IoT nodes in our problem formulation.

\subsection{Information Coding for Transmission}

Let us consider the $j$th IoT node communicates with the $i$th gateway separated by Euclidean distance $d_{ji}$. Let the $j$th node fire with transmit power $\mathcal{P}_T(j)$ and use an encoding function $\mathsf{f}(j)$ to generate a sequence of $\bar{\mathsf{n}}$ input symbols from a sequence of random variables $\tilde{\mathbb{W}}$ such that, $\tilde{\mathbb{W}} = \mathsf{f}(\tilde{\mathcal{X}}^j_1, \dotso, \tilde{\mathcal{X}}^j_{\bar{\mathsf{n}}})$ is the input on the link between $(j,i)$. For the $i$th gateway, let the $\bar{\mathsf{k}}$th received symbol be $\tilde{\mathcal{Y}}^i_{\bar{\mathsf{k}}}$, so we can write,
\begin{align}\label{eq19}
    \text{Pr}\{\tilde{\mathcal{Y}}^i_{\bar{\mathsf{k}}} = \tilde{\mathsf{y}}|\tilde{\mathcal{X}}^j_{\bar{\mathsf{k}}} = \tilde{\mathsf{x}}, \varepsilon_{\bar{\mathsf{k}}}\} = \mathbb{Q}_{ji}(\tilde{\mathsf{y}}|\tilde{\mathsf{x}})
\end{align}
where $\tilde{\mathcal{X}}$ represents the input alphabet and $\tilde{\mathcal{Y}}$ represents the output alphabet of the links between the IoT end-nodes and gateways, $\mathbb{Q}$ is used to represent the actual link as a transfer function between the input and the output, and $\varepsilon_{\bar{\mathsf{k}}}$ is a set of random variables. Let the decoding function of the $i$th gateway be given by $\mathsf{g}$ and the recovered message be of the form $\tilde{\mathbb{W}}'$ denoted by, $\tilde{\mathbb{W}}' = \mathsf{g}(\tilde{\mathcal{Y}}^i_1, \dotso, \tilde{\mathcal{Y}}^i_{\bar{\mathsf{n}}})$, where $\tilde{\mathbb{W}}'$ is also a random variable of length $\bar{\mathsf{m}}$. Then the error probability performance of the link is given by,
\begin{align}\label{eq20}
    \Phi = \max_{\tilde{\mathbb{W}} \in \{1, \dotso, \bar{\mathsf{m}}\}} \text{Pr}\{\tilde{\mathbb{W}}' \neq \tilde{\mathbb{W}}\}
\end{align}
where, $\Phi$ is the maximum probability of error in recovering the original message $\tilde{\mathbb{W}}$, considering all possible original messages within the set $\{1, \dotso, \bar{\mathsf{m}}\}$ and a coding rate ${\mathcal{R}}$ is achievable if there exists a sequence of $(\lceil 2^{\bar{\mathsf{n}},\mathcal{R}}\rceil, \bar{\mathsf{n}})$ codes such that $\Phi$ tends to 0 as $\bar{\mathsf{n}}$ becomes infinitely large. In turn, we can also calculate the capacity of the link which is the supremum of all the achievable rates.

\vspace{-5mm}
\subsection{Transmit Power Allocation}

If we consider $\mathcal{P}_T(j)$ is the transmit power of the $j$th node, the received power after the $j$th node fires its signal can be expressed as,
\begin{align}\label{eq21}
   \mathcal{P}_T(\mathsf{r}_j) = \kappa_1 \mathsf{r}^{\nu}_j + \kappa_2
\end{align}
where $\mathsf{r}_j$ is the transmission range of the $j$th node, $\nu$ is the pathloss exponent, $\kappa_1$ is the constant defining the IoT node type, like whether it is a bi-static or a multi-static sensor, and $\kappa_2$ quantifies the link quality between the $j$th node and the $i$th gateway. Now on the gateway side, the received power can be calculated as,
\begin{align}\label{eq22}
   \mathcal{P}^{ij}_R(d_{ji}) = \frac{\mathcal{P}^{ij}_R(d_{0i}) \times d^{\nu}_{0i}}{d^{\nu}_{ji}}
\end{align}
where $\mathcal{P}^{ij}_R(d_{ji})$ is the received power at the $i$th gateway from the $j$th node over the Euclidean distance $d_{ji}$ and $\mathcal{P}^{ij}_R(d_{0i})$ is the received power at the $i$th gateway over the far-field distance $d_{0i}$. Using the above formulations in (\ref{eq21}) and (\ref{eq22}), we can design the error probability minimization problem, with constraint on code rate maximization as,
\begin{align}\label{eq23}
\text{minimize}_{\mathbb{Q}_{ji}}~~\mathcal{G}(\Phi) &= \Bigg[\frac{\mathsf{r}^{\nu}_j}{\mathcal{E}_j} + d^{\nu}_{ji}\Bigg]\times \mathbb{Q}_{ji} \nonumber\\
\text{subject to,}~~& \mathbb{Q}_{ji} = \max_{a \in V} (\mathsf{h}_{ja})~\text{and}~0 \leq \mathsf{h}_{ja} \leq \mathcal{R}_{ja}\nonumber\\
&\text{and}~0 < \mathsf{r}_j \leq \mathsf{r}_{\text{max}}\nonumber\\
&\sum_{a|a \in V} \mathsf{h}_{ja} - \sum_{a|j \in V} \mathsf{h}_{aj} = \psi_j~\text{for}~d_{ji} \leq \mathsf{r}_j
\end{align}
where $\mathcal{E}_j$ is the remaining energy at the $j$th node after transmission, $\mathsf{h}_{ji}$ is the virtual flow rate of symbols on the link $(j,i)$, $\mathcal{R}_{ja}$ is maximum achievable code rate over the link between the $j$th and $a$th nodes, $\mathsf{r}_{\text{max}}$ is the average maximum transmission range of any node within the network, and $\mathbb{Q}_{ji}$ is the link transfer function between the $j$th node and the $i$th gateway. Details on the optimality of (\ref{eq23}) is provided in Appendix~C.

\subsection{Topology Extraction}

Next we will be using the solution to the code design problem formulated in (\ref{eq23}), and improve information or code rate. First we compute the transmit power assignment to minimize error probability and then extract the corresponding network topology. The topology extraction algorithm continues iteratively until it converges to a point where error probability is minimum and code rate is maximum. The extraction algorithm is presented in Algorithm 2\footnote{The terms "sew" and "stitch" are used interchangeably to describe the process of combining or joining smaller sub-graphs to form a larger, unified graph representation of the IoT network.}.\\
\vspace{-4mm}
\begin{algorithm}[h]
\caption{~~Topology Extraction for Minimizing Error Probability (\textbf{IoTNTop})}\label{tab:x}
\textbf{Input} : A set of IoT nodes in $V$ and their coordinates $\{X,Y\}$ and the iteration step of small value $\varsigma$\\
\textbf{Step 1} : Power assignment $\mathcal{P}_T(j) \to \{\mathcal{P}_T(1), \dotso, \mathcal{P}_T(l)\}$. For all node pairs $i,j$ such that $d_{ij} \leq$ transmission range, compute SNR.\\ 
\textbf{Step 2} : Sort edges in the decreasing order of SNR\\
Let $\hat{e}_1, \hat{e}_2, \dotso$ be the resulting sequence of edges, initialize $m$ patches\\
If patch ($i$) $\neq$ patch ($j$) then sew them using graph embedding algorithm outlined in Section II, $E = E \cup \{\hat{e}_{i,j}\}$, where $\hat{e}_{i,j}$ is the newly formed edge after patch alignment\\
\textbf{Extract Topology} $\Gamma(V,E)$ \\
\textbf{Step 3} : Calculate SNR$(\Gamma, \mathsf{P}_T)$ and initialize $\Delta = 1$\\
For all; $\Delta > \varepsilon$; SNR$_{\text{max}}$ = SNR$(\Gamma, \rho_{T_{\text{max}}})$; $\Delta = ||\text{SNR}_{\text{max}} - \text{SNR}(\Gamma, \mathsf{P}_T)||$\\
\textbf{Output} : Finalize power assignment $\mathsf{P}_T$
\end{algorithm}
\vspace{-5mm}
\subsection{Convergence Analysis}

The overall optimization problem of minimizing error probability and maximizing code-rate within the network topology is non-convex. This makes it very challenging to perform direct convergence analysis by using techniques like contraction mappings and fixed-point theorems. Therefore, we resort to applying a Lyapunov Function approach. Since the objective function $\mathcal{G}(\Phi)$ mixes error probability, energy and geometric quantities like distances and ranges, the formulated Lyapunov function will capture the combined behavior of these intertwined variables. We consider the global error probability metric ($\Phi$) directly as the potential Lyapunov function; $\mathcal{V}(\tilde{x}) = \Phi$ where $\tilde{x}$ is the state of the network including power allocations. To include topological measure, we modify to, $\mathcal{V}(\tilde{x}) = \Phi + \beta*\mathcal{M}(\text{top})$
where $\mathcal{M}(\text{top})$ is the topology metric and $\beta$ is the weighting factor to adjust the influence of the topology. We define $\mathcal{M}(\text{top})$ as,
\begin{align}
    \mathcal{M}(\text{top}) = (1/N_{\text{links}})\sum_{j,i} \mathbb{Q}_{ji}
\end{align}
where $N_{\text{links}}$ is the total number of active links in the network and $\mathbb{Q}_{ji}$ is the quality of the link between $j$th node and $i$th gateway. $\mathbb{Q}_{ji}$ can be defined as the received signal-to-noise ratio (SNR) of a reliably detectable signal formulated as,
\begin{align}
    \mathbb{Q}_{ji} = h_{ji}\mathcal{P}_T(j) d_{ji}^{-\nu}/\mathcal{N} \geq \zeta
\end{align}
where $h_{ji}$ is the link gain between $j$th node and $i$th gateway, $d_{ji}$ is the Euclidean distance between them, $\nu$ is the pathloss exponent, $\mathcal{N}$ is the noise power and $\zeta$ is the lower SNR threshold or the minimum detectable SNR on the receive side. In order to check the convergence of the algorithm, we need to continue as long as $\mathcal{V}(\tilde{x})$ decreases along the trajectory of the system. Therefore, IoTNTop continues to iterate as long as $\mathcal{V}(\tilde{x}_{k+1}) - \mathcal{V}(\tilde{x}_{k}) < 0$ where $k$ represents the number of iterations.
\vspace{-5mm}

\subsection{Initial Power Assignment}

We also need to think about the initial power assignment to the IoT nodes at the beginning of IoTNTop. Here we explore distance-based power assignment, where the initial transmit power is tailored to the distance between a node and its nearest gateway. Let us denote the distance between all IoT nodes and their nearest gateways by using a distance matrix $\mathcal{D}$ of size $l \times s$, where $l$ and $s$ are the number of nodes and gateways, and $d_{ji}$ is an element of $\mathcal{D}$. We incorporate a pathloss model to account for signal attenuation with distance.

A common model is the Friis transmission equation, $\text{PL(dB)} = 10\nu \log_{10}(d_{ji}) + \tilde{\Gamma}$, where $\tilde{\Gamma}$ is a constant depending on environment and node properties. We want to find the minimum initial transmit power to ensure the signal reaches the receiver with sufficient strength for successful reception, while considering the path loss. Each receiver has a sensitivity threshold, which is the minimum power level it needs to correctly decode the signal. We assume we know this value. We solve the Friis formula for received power given a transmit power, set the received power to the minimum required level and solve this equation for the initial transmit power, $\mathcal{P}_{T\text{init, i}}$. If $\mathcal{P}_{T\text{min}}$ is the minimum transmit power required for a successful transmission under ideal conditions, we can calculate the initial power, $\mathcal{P}_{T\text{init, i}}$ as, $\mathcal{P}_{T\text{init, i}} = \mathcal{P}_{T\text{min}}*10^{\text{PL(dB)}/10 (d_{j, \text{nearest gateway} (j)})}$, where $\text{nearest gateway} (j)$ is a function that returns the index of the gateway closest to the $j$th node and the propagation pathloss is calculated using the pathloss exponent between the $j$th node and its nearest gateway.
\begin{figure}[t]
\begin{center}
    \includegraphics[width=0.9\columnwidth]{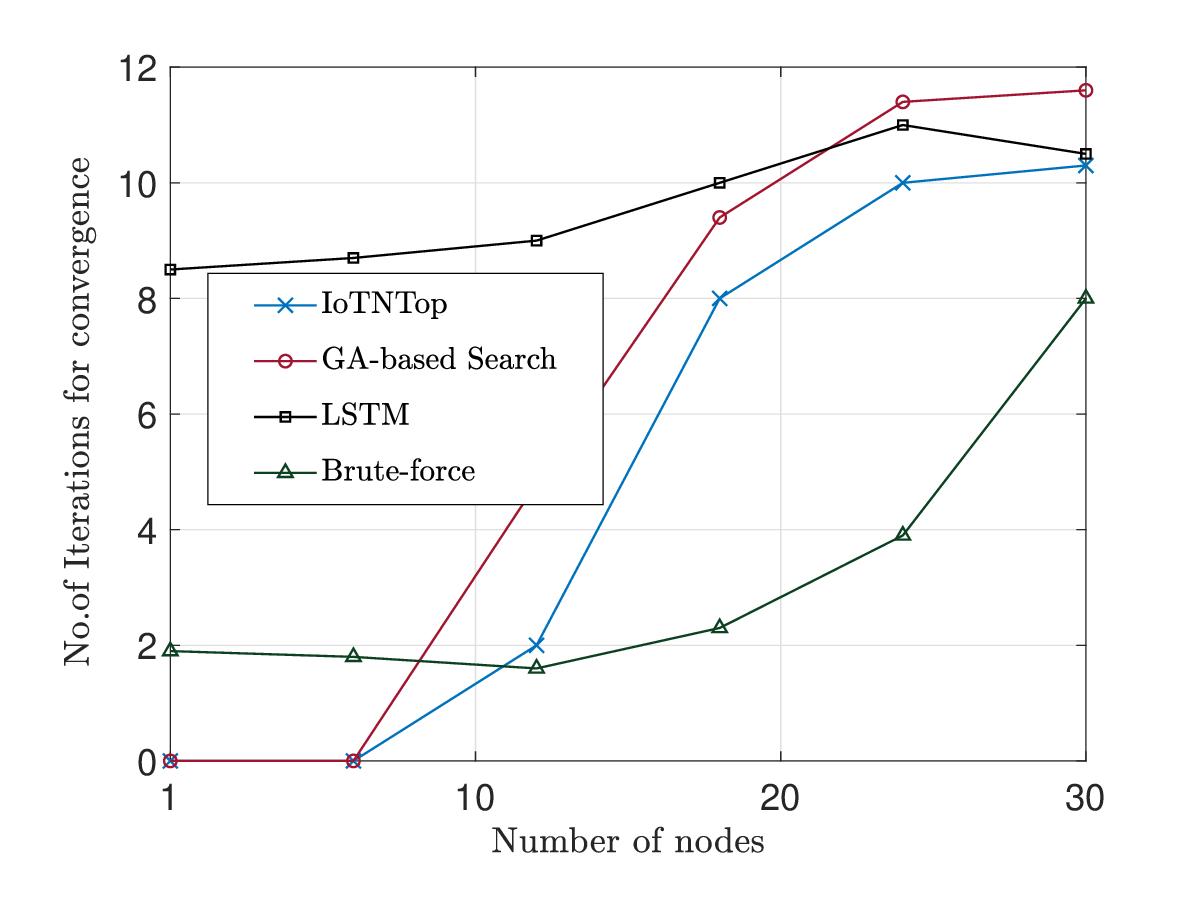}
\end{center}
\vspace{-5mm}
\caption{Variation in the number of iterations needed by different topology control algorithms as the number of nodes within the network increases; the nodes are considered to be randomly distributed over a 5 $\times$ 5 km$^2$ square area.}
\vspace{-8mm}
\label{FIG2}
\end{figure}

\section{Numerical Results and Discussion} 

We compare the results of our proposed \textbf{IoTNTtop} algorithm with those of Brute-force search, by choosing the minimum transmit power allocation to any IoT node for firing its signal to its farthest neighboring node in the graph. The Brute-force algorithm will systematically enumerate all the possible solutions to the transmit power allocation problem in (\ref{eq23}), until all optimal solutions have been exhausted. Our proposed algorithm is also compared with the local minimal spanning tree (LMST) \cite{25} which is a topology control algorithm for communication graphs. We focus on a non-link-sharing scenario, where each link is used only between a pair of nodes in a particular iteration. We consider 100 IoT nodes and 6 gateways distributed randomly within a $4\times 4$ km$^2$ square area without any central gateway. We considered average pathloss exponent of $\nu = 2$, $\rho_{R_{\text{min}}} = -63$ dBm, $\rho_{T_{\text{max}}} = 27$ dBm, $\beta = 2.5$ and a typical value of $\mathcal{N} = - 50$ dBm is considered for all simulation results. $\rho_{R_{\text{min}}} = -63$ dBm represents the minimum received power level required at the gateway for successful decoding of the signal. Any signal received with power below this threshold is considered too weak to be reliably interpreted. $\rho_{T_{\text{max}}} = 27$ dBm represents the maximum transmit power level that an IoT node is allowed to use. This constraint is chosen due to hardware capabilities of the nodes with a desire to limit interference in the network.

Fig.~\ref{FIG2} represents the convergence behavior of different algorithms compared to IoTNTop against the number of IoT nodes within the network. IoTNTop's efficiency in terms of network resource utilization helps it converge with fewer number of nodes as compared to other algorithms. The Brute-Force curve starts at a higher number of nodes and slowly decreases. This is because a brute-force search exhaustively evaluates all possible configurations, initially considering a large number of nodes before converging to a minimum.  A genetic Algorithm (GA)-based approach also explores many possibilities before converging. IoTNTop optimizing network topology, using fewer nodes implies a more streamlined network structure that reduces redundancy or unnecessary connections. Furthermore, as IoTNTop is optimizing power control, it uses fewer active nodes, effectively putting some nodes into sleep mode to conserve energy.

\begin{figure}[t]
\begin{center}
    \includegraphics[width=0.9\columnwidth]{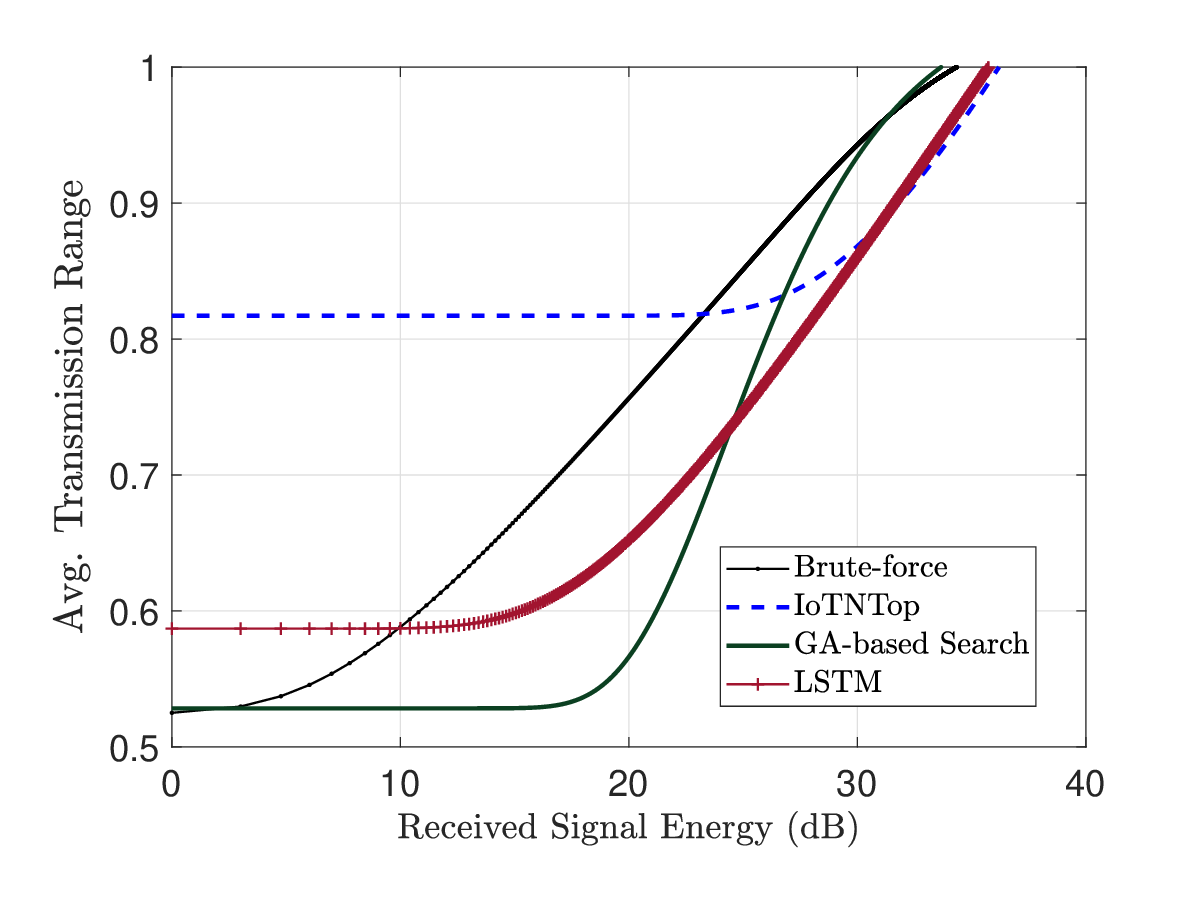}
\end{center}
\vspace{-5mm}
\caption{Comparative variation in average transmission range per node within the network, over different values of average received signal energy per node in an IoT network consisting of 100 nodes randomly distributed over a 5 × 5 km$^2$ square area.}
\vspace{-8mm}
\label{FIG3}
\end{figure}

Fig.~\ref{FIG3} shows the average transmission range (in meters) achieved by different algorithms over a course of iterations. IoTNTop demonstrates superior efficiency in achieving a stable average transmission range, suggesting a selective approach to finding suitable network configurations. While the Brute-Force method eventually reaches a similar range, its slower convergence highlights its exhaustive exploration of all possibilities. GA displays a more explorative nature, with fluctuations before settling on a lower average transmission range, indicating it considers a wider variety of solutions. In contrast, LSTM prioritizes rapid convergence, but its lower average transmission range suggests it might not always find the absolute optimal solution. It is possible that the LSTM got stuck in a local optimum, or it might be prioritizing other factors besides maximizing transmission range.

\begin{figure}[t]
\begin{center}
    \includegraphics[width=0.8\columnwidth]{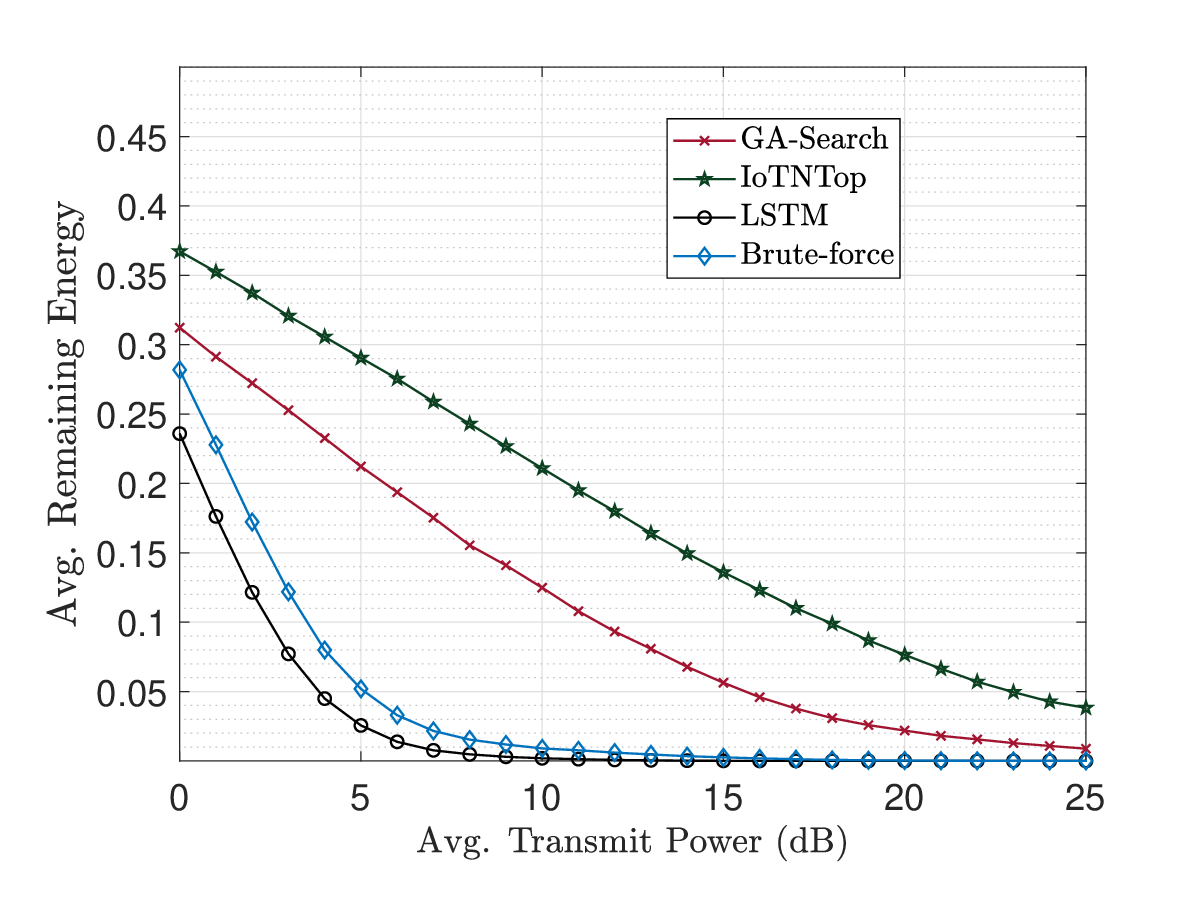}
\end{center}
\vspace{-5mm}
\caption{Comparative variation in average energy supply remaining in each node within the network over different values of average transmit signal power per node in an IoT network consisting of 100 nodes randomly distributed over a 5 × 5 km$^2$ square area.}
\vspace{-8mm}
\label{FIG4}
\end{figure}

\begin{figure}[t]
\begin{center}
    \includegraphics[width=0.9\columnwidth]{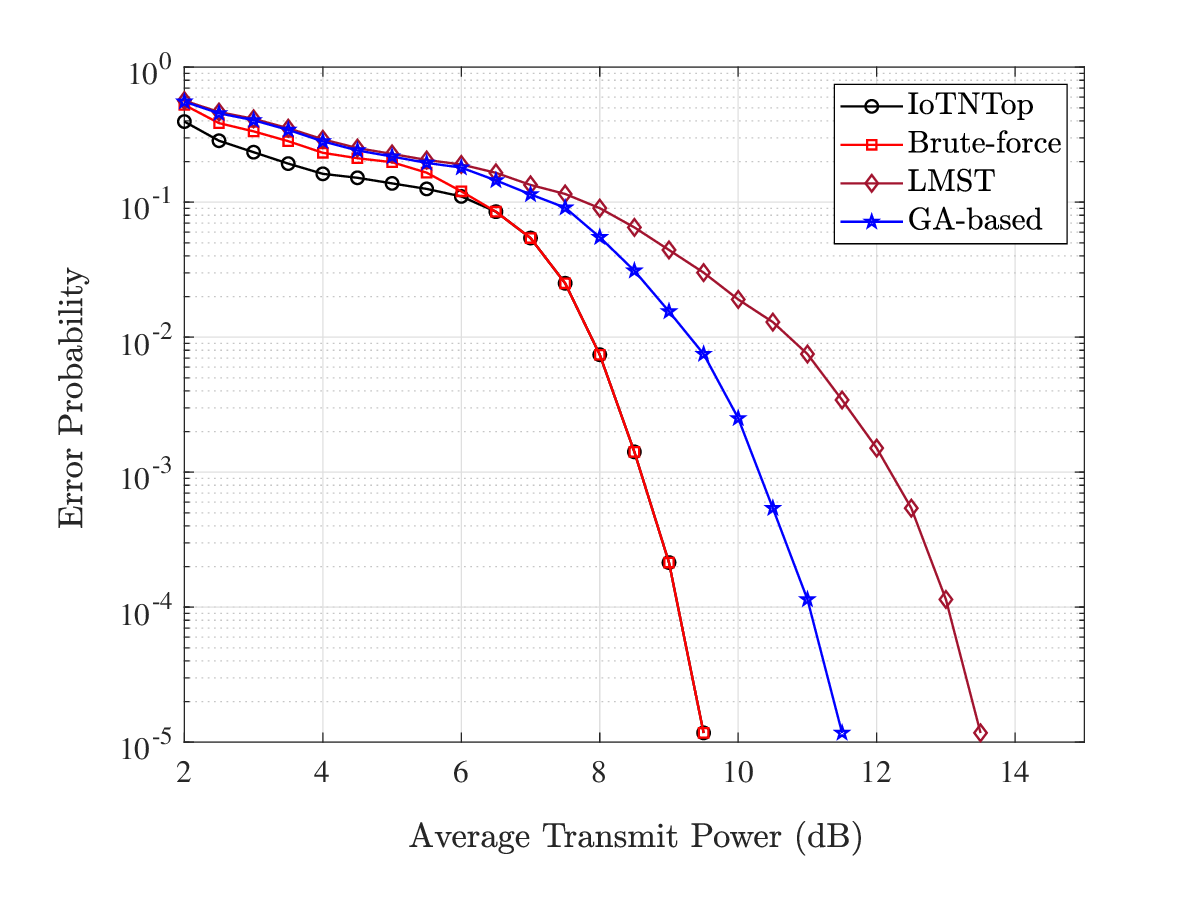}
\end{center}
\vspace{-5mm}
\caption{Comparative variation in average error probability in transmission per node within the network, over different values of average transmit signal power per node in an IoT network consisting of 100 nodes randomly distributed over a 5 × 5 km$^2$ square area.}
\vspace{-5mm}
\label{FIG5}
\end{figure}

\begin{figure}[t]
\begin{center}
    \includegraphics[width=0.9\columnwidth]{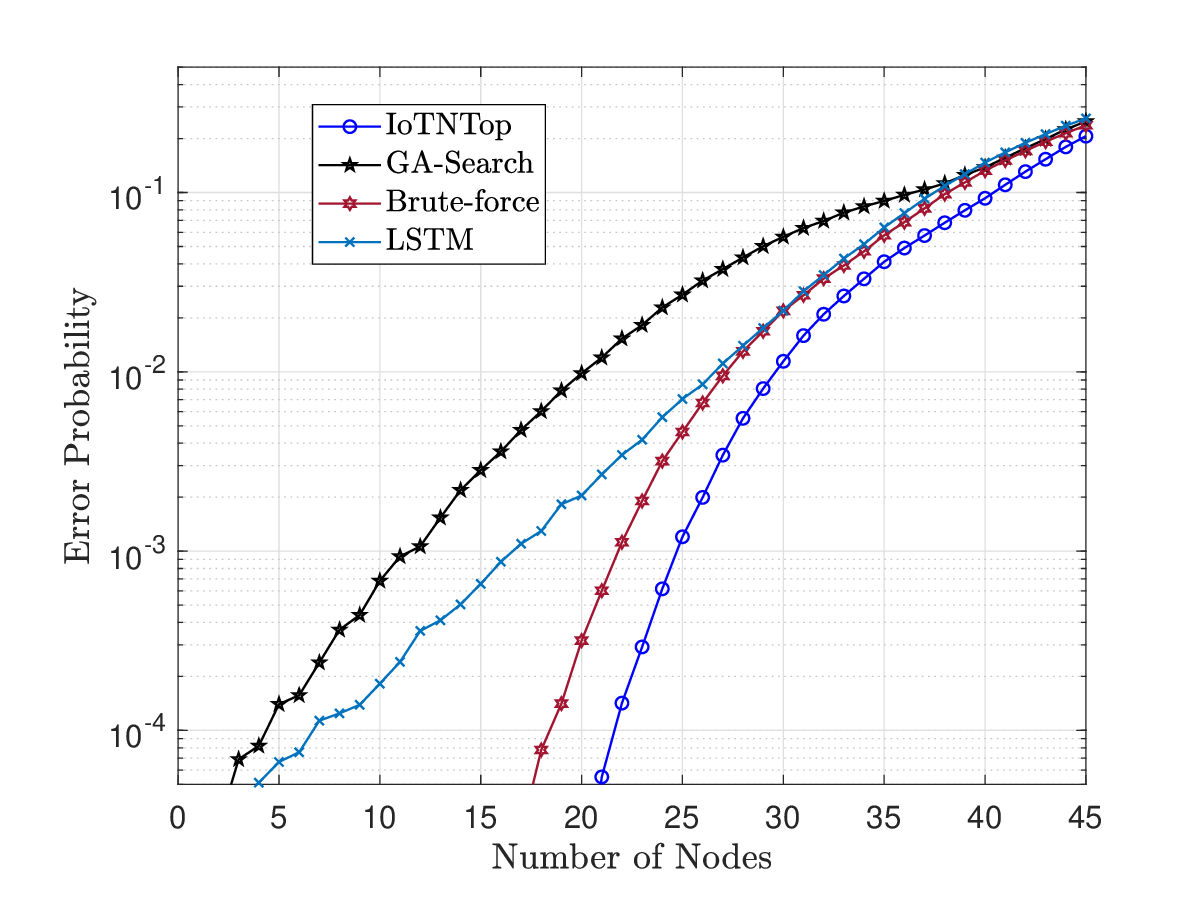}
\end{center}
\vspace{-5mm}
\caption{Comparative variation in average error probability in transmission per node within the network as the number of nodes within the network increases; the nodes are considered to be randomly distributed over a 5 × 5 km$^2$ square area.}
\vspace{-5mm}
\label{FIG6}
\end{figure}

Compared to the other algorithms, IoTNTop consistently achieves a lower average transmit power throughout the iterations, as shown in Fig.~\ref{FIG4}. This reduction in transmit power translates to several benefits for the network.  First, nodes will consume less energy to transmit data, which can significantly extend battery life in resource-constrained IoT devices. Battery life is a critical factor for the successful deployment of large-scale IoT networks, especially in applications where frequent battery replacements are impractical or infeasible. Second, lower power transmissions can help mitigate interference between nodes. This is because the strength of a signal attenuates with distance, and weaker signals are more susceptible to interference from overlapping transmissions. By reducing the transmit power, IoTNTop helps ensure that signals from different nodes are less likely to interfere with each other, improving overall network reliability and throughput. This can be especially important in dense IoT deployments where many devices are competing for limited channel bandwidth.

IoTNTop's relatively fast convergence to a stable average transmit power suggests it adopts an efficient strategy for searching the solution space. In contrast, Brute-Force, while achieving a similar power level eventually, requires a significantly longer number of iterations due to its exhaustive exploration of all possible configurations. This brute-force approach guarantees finding the optimal solution but comes at the high cost of computational complexity. The GA and LSTM curves, on the other hand, exhibit more fluctuations during their convergence process. This indicates that these algorithms are employing more explorative search strategies, evaluating a wider range of candidate solutions before settling on a final configuration. This exploration comes with the benefit of potentially uncovering alternative solutions that are not identified by a more directed search, but it will also lead to a slower convergence rate compared to IoTNTop.
\begin{figure}[t]
\begin{center}
    \includegraphics[width=0.9\columnwidth]{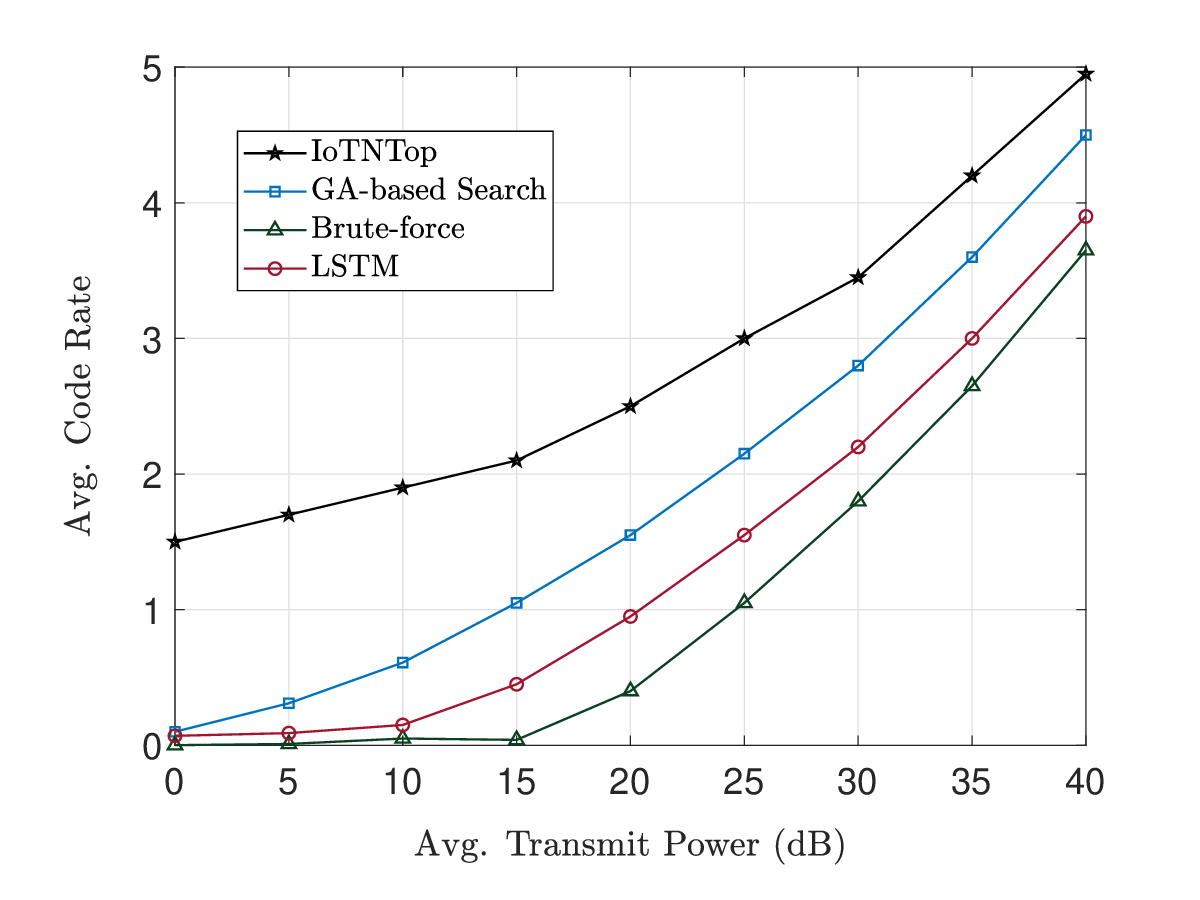}
\end{center}
\vspace{-5mm}
\caption{Comparative variation in achievable transmission code rate per node within the network over different values of average transmit signal power per node in an IoT network consisting of 100 nodes randomly distributed over a 5 × 5 km$^2$ square area.}
\vspace{-5mm}
\label{FIG7}
\end{figure}

IoTNTop's error rate curve in Fig.~\ref{FIG5} shows steady improvement, likely due to its ability to optimize configurations that reduce transmit power and improve transmission quality. While Brute-Force achieves a similar low error rate, it requires significantly more computational effort. The GA and LSTM algorithms also reduce error rates, but their broader exploration of configurations slows convergence, with GA showing a gradual decrease and LSTM exhibiting fluctuations. Although exploring a wider solution space can yield diverse results, it delays reaching optimal configurations compared to IoTNTop’s more efficient approach.

\begin{figure}[t]
\begin{center}
    \includegraphics[width=0.9\columnwidth]{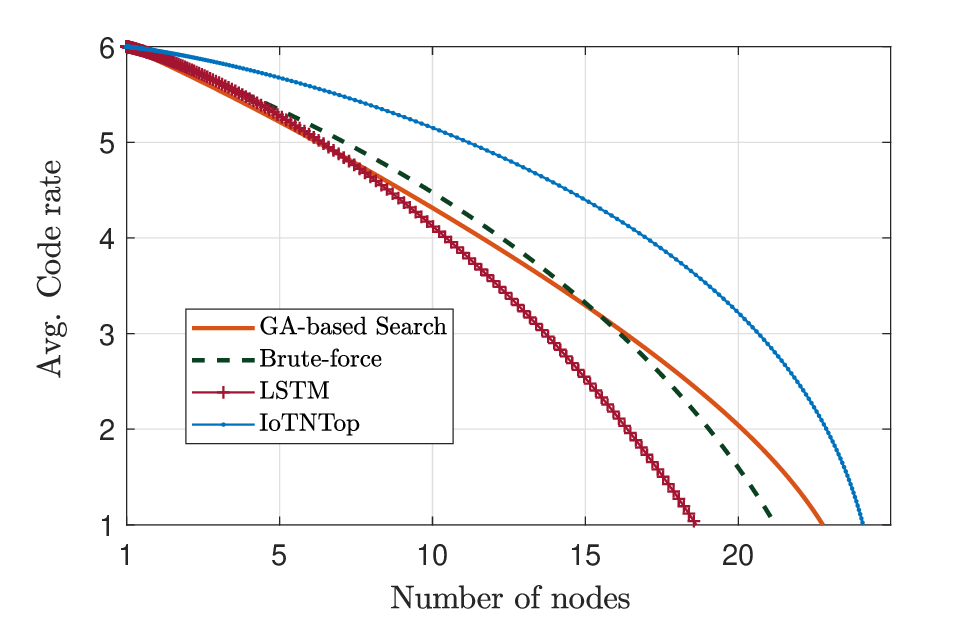}
\end{center}
\vspace{-3mm}
\caption{Comparative variation in achievable transmission code rate per node within the network as the number of nodes within the network increases; the nodes are considered to be randomly distributed over a 5 × 5 km$^2$ square area.}
\vspace{-8mm}
\label{FIG8}
\end{figure}

Compared to the other algorithms, IoTNTop achieves a significantly higher transmission code rate throughout the iterations (refer to Fig.~\ref{FIG7} and Fig.~\ref{FIG8}). Higher code rates mean fewer redundant bits, reducing errors and improving packet delivery rates in noisy channels. While Brute-Force also achieves high code rates, it requires more iterations, indicating IoTNTop’s more efficient, targeted approach. The GA and LSTM algorithms show a slower increase in code rate, suggesting broader exploration before convergence. Overall, IoTNTop's ability to quickly optimize code rates is key to enhancing reliable data transmission and reducing packet loss.
\vspace{-3mm}
\section{Conclusions}

This paper tackles the challenge of designing IoT network topologies to maximize coverage and throughput with minimal number of devices while addressing issues like poor link quality and interference. By introducing a novel graph-realization approach that incorporates local distance information into the global network structure, the authors optimize the placement of end-nodes and gateways, improving transmit power, information rates, and reducing error probability. The methodology divides the network into sub-graphs, aligning them to form an efficient global structure. Through advanced mathematical modeling and algorithms, this scalable, noise-resilient solution outperforms traditional methods in minimizing errors and maximizing code rates, making it ideal for large, heterogeneous IoT networks.
\vspace{-5mm}






\begin{appendices}
\section{Optamility of (\ref{eq4c})}

\noindent a) \emph{Objective Function Analysis}: The objective function is $\sum_{\bar{i} = 1}^m ||\bar{\mathsf{s}}e^{\imath\theta} (\bar{x}_i - \mu_{\bar{X}}) + (\bar{t} - \mu_{\bar{Y}}) - (\bar{y}_i - \mu_{\bar{Y}})||^2$. This function measures the squared Euclidean distance between the transformed \(\bar{x}_i\) points and the \(\bar{y}_i\) points.

\noindent b) \emph{Define the Transformation}: Let \(\mathsf{s}\) be the scaling factor, \(\theta\) be the rotation angle, and \(\bar{t}\) be the translation vector. The transformation applied to \(\bar{x}_i\) is \(\bar{\mathsf{s}}e^{\imath\theta} (\bar{x}_i - \mu_{\bar{X}}) + (\bar{t} - \mu_{\bar{Y}})\).

\noindent c) \emph{Optimality Conditions}: To find the optimal \(\mathsf{s}\), \(\theta\), and \(\bar{t}\), we need to minimize the sum of squared differences. We achieve this by finding the values that minimize the distance between the transformed points and the target points.

\noindent d) \emph{Derivation of Optimal \(\bar{t}\)}: The optimal translation \(\bar{t}\) aligns the centroids of the transformed \(\bar{x}_i\) points and the \(\bar{y}_i\) points. The centroids of the sets are \(\mu_{\bar{X}}\) and \(\mu_{\bar{Y}}\). Therefore, the optimal translation is $\bar{t} = \mu_{\bar{Y}} - \mathsf{s}e^{\imath\theta}\mu_{\bar{X}}$.

\noindent e) \emph{Derivation of Optimal \(\mathsf{s}\) and \(\theta\)}: The optimal scaling and rotation can be derived using Singular Value Decomposition (SVD) of the centered data matrices. Let \(X_c\) and \(Y_c\) be the centered data matrices:
\begin{align}
    X_c &= [\bar{x}_1 - \mu_{\bar{X}}, \bar{x}_2 - \mu_{\bar{X}}, \ldots, \bar{x}_m - \mu_{\bar{X}}]\\
    Y_c &= [\bar{y}_1 - \mu_{\bar{Y}}, \bar{y}_2 - \mu_{\bar{Y}}, \ldots, \bar{y}_m - \mu_{\bar{Y}}]
\end{align}
The cross-covariance matrix \(H\) is: $H = X_c Y_c^T$. Performing SVD on \(H\): $ H = U \Sigma V^T.$ The optimal rotation matrix \(R\) is given by: $R = V U^T$. The optimal scaling factor \(\mathsf{s}\) is $\mathsf{s} = \frac{\text{trace}(Y_c^T R X_c)}{\text{trace}(X_c^T X_c)}$. Final Optimal Values include,
\begin{itemize}
    \item The optimal rotation angle \(\theta\) is derived from the rotation matrix \(R\): $e^{\imath\theta} = R$
    \item The optimal scaling factor is \(\mathsf{s}\): $\mathsf{s} = \frac{\text{trace}(Y_c^T R X_c)}{\text{trace}(X_c^T X_c)}$
    \item The optimal translation vector \(\bar{t}\) is: $\bar{t} = \mu_{\bar{Y}} - \mathsf{s}e^{\imath\theta}\mu_{\bar{X}}$
\end{itemize}
By aligning the centroids, applying the optimal rotation, and scaling, we minimize the objective function. The optimal transformation parameters \(\mathsf{s}\), \(e^{\imath\theta}\), and \(\bar{t}\) achieve the minimum sum of squared differences, proving the optimality of the given problem.
\vspace{-4mm}

\section{Optimality of (\ref{eq8})}

\noindent Let \(\mathbf{0}\) be a zero vector of appropriate dimension and \(e_i\) and \(e_j\) be the \(i\)-th and \(j\)-th standard basis vectors, respectively. Then:
\begin{align}
    (\mathbf{0}; e_i - e_j) = \begin{bmatrix} \mathbf{0} \\ e_i - e_j \end{bmatrix}
\end{align}
Consequently:
\begin{align}
    (\mathbf{0}; e_i - e_j)(\mathbf{0}; e_i - e_j)^T = \begin{bmatrix} \mathbf{0} & \mathbf{0} \\ \mathbf{0} & e_i - e_j \end{bmatrix} \begin{bmatrix} \mathbf{0} & \mathbf{0} \\ \mathbf{0} & e_i - e_j \end{bmatrix}^T
\end{align}
This simplifies to:
\begin{align}
    \begin{bmatrix} \mathbf{0} & \mathbf{0} \\ \mathbf{0} & (e_i - e_j)(e_i - e_j)^T \end{bmatrix}
\end{align}
The term \((e_i - e_j)(e_i - e_j)^T\) is a rank-1 matrix with non-zero elements only in the \(i\)-th and \(j\)-th positions. The constraint \((\mathbf{0}; e_i - e_j)(\mathbf{0}; e_i - e_j)^T \chi = d^2_{ij}\) implies that \(\chi\) must be such that when pre-multiplied and post-multiplied by \((\mathbf{0}; e_i - e_j)\), the result is \(d^2_{ij}\). This enforces a specific structure on \(\chi\). Specifically: $(\mathbf{0}; e_i - e_j)^T \chi (\mathbf{0}; e_i - e_j) = d^2_{ij}$. The matrix equation simplifies to $(e_i - e_j)^T \chi (e_i - e_j) = d^2_{ij}$. Since \(e_i\) and \(e_j\) are standard basis vectors, \(\chi\) must have the property that $\chi_{ii} - 2\chi_{ij} + \chi_{jj} = d^2_{ij}$. This equation ensures that the squared distance \(d^2_{ij}\) between the \(i\)-th and \(j\)-th positions is maintained. Given that the objective function is a constant zero, the problem reduces to finding \(\chi\) such that the constraint \((e_i - e_j)^T \chi (e_i - e_j) = d^2_{ij}\) is satisfied. This constraint enforces that the matrix \(\chi\) correctly represents the distances \(d_{ij}\) between points \(i\) and \(j\). The constraints are linear in nature and define a feasible region. Since the objective function is zero and does not change, any \(\chi\) that satisfies the constraint is optimal. Therefore, the problem is optimal because the constraints alone define the feasible set, and any \(\chi\) that satisfies the constraint will achieve the objective function value of zero, proving the optimality.

\section{Optamility of (\ref{eq23})}
To find the optimal \(\mathbb{Q}_{ji}\) and \(\mathsf{h}_{ja}\), let's solve the optimization problem step-by-step:
The optimization problem is:
\begin{align}
   \text{minimize}_{\mathbb{Q}_{ji}}~~\mathcal{G}(\Phi) &= \Bigg[\frac{\mathsf{r}^{\nu}_j}{\mathcal{E}_j} + d^{\nu}_{ji}\Bigg]\times \mathbb{Q}_{ji} \\
   \text{subject to,}~~& \mathbb{Q}_{ji} = \max_{a \in V} (\mathsf{h}_{ja}) \nonumber\\
   & 0 \leq \mathsf{h}_{ja} \leq \mathcal{R}_{ja}, 0 < \mathsf{r}_j \leq \mathsf{r}_{\text{max}} \nonumber\\
   & \sum_{a|a \in V} \mathsf{h}_{ja} - \sum_{a|j \in V} \mathsf{h}_{aj} = \psi_j~\text{for}~d_{ji} \leq \mathsf{r}_j \nonumber
\end{align}
The Lagrangian is given by:
\begin{align}
   \mathcal{L} &= \Bigg[\frac{\mathsf{r}^{\nu}_j}{\mathcal{E}_j} + d^{\nu}_{ji}\Bigg]\times \mathbb{Q}_{ji} + \lambda_1 (\mathbb{Q}_{ji} - \max_{a \in V} (\mathsf{h}_{ja})) \nonumber\\
   &+ \sum_{a \in V} \lambda_{2a} (\mathsf{h}_{ja} - \mathcal{R}_{ja}) + \sum_{a \in V} \lambda_{3a} (-\mathsf{h}_{ja}) + \lambda_4 (\mathsf{r}_j - \mathsf{r}_{\text{max}}) \nonumber\\
   &+ \lambda_5 (-\mathsf{r}_j) + \lambda_6 \left(\sum_{a \in V} \mathsf{h}_{ja} - \sum_{a \in V} \mathsf{h}_{aj} - \psi_j\right)
\end{align}
Using the Karush-Kuhn-Tucker (KKT) conditions we formulate;\\
a) \emph{Stationarity}:
\begin{align}
    \frac{\partial \mathcal{L}}{\partial \mathbb{Q}_{ji}} = \Bigg[\frac{\mathsf{r}^{\nu}_j}{\mathcal{E}_j} + d^{\nu}_{ji}\Bigg] + \lambda_1 = 0 \quad \Rightarrow \quad \lambda_1 = -\Bigg[\frac{\mathsf{r}^{\nu}_j}{\mathcal{E}_j} + d^{\nu}_{ji}\Bigg]
\end{align}
\begin{align}
    \frac{\partial \mathcal{L}}{\partial \mathsf{h}_{ja}} = -\lambda_1 \frac{\partial (\max_{a \in V} (\mathsf{h}_{ja}))}{\partial \mathsf{h}_{ja}} + \lambda_{2a} - \lambda_{3a} + \lambda_6 = 0 
\end{align}
Note: The partial derivative \(\frac{\partial (\max_{a \in V} (\mathsf{h}_{ja}))}{\partial \mathsf{h}_{ja}}\) is 1 if \(\mathsf{h}_{ja} = \max_{a \in V} (\mathsf{h}_{ja})\) and 0 otherwise.
\begin{align}
    \frac{\partial \mathcal{L}}{\partial \mathsf{r}_j} = \frac{\nu \mathsf{r}^{\nu-1}_j}{\mathcal{E}_j} \mathbb{Q}_{ji} - \lambda_4 + \lambda_5 = 0 \quad \Rightarrow \quad \lambda_4 - \lambda_5 = \frac{\nu \mathsf{r}^{\nu-1}_j}{\mathcal{E}_j} \mathbb{Q}_{ji}
\end{align}
b) \emph{Primal Feasibility}:\\
The original constraints should hold:
\begin{itemize}
    \item \(\mathbb{Q}_{ji} = \max_{a \in V} (\mathsf{h}_{ja})\)
    \item \(0 \leq \mathsf{h}_{ja} \leq \mathcal{R}_{ja}\)
    \item \(0 < \mathsf{r}_j \leq \mathsf{r}_{\text{max}}\)
    \item \(\sum_{a \in V} \mathsf{h}_{ja} - \sum_{a \in V} \mathsf{h}_{aj} = \psi_j\) for \(d_{ji} \leq \mathsf{r}_j\)
\end{itemize}
c) \emph{Dual Feasibility}: \(\lambda_i \geq 0\) for all \(i\).\\
d) \emph{Complementary Slackness}: Each \(\lambda_i\) should be zero if the corresponding constraint is strictly less than its bound.\\
e) \emph{Solving for Optimal Values}:
\begin{itemize}
\item From the stationarity condition with respect to \(\mathbb{Q}_{ji}\): 
\begin{align}
    \lambda_1 = -\Bigg[\frac{\mathsf{r}^{\nu}_j}{\mathcal{E}_j} + d^{\nu}_{ji}\Bigg]
\end{align}
\item From the stationarity condition with respect to \(\mathsf{h}_{ja}\): \\
If \(\mathsf{h}_{ja} = \max_{a \in V} (\mathsf{h}_{ja})\): $-\lambda_1 + \lambda_{2a} - \lambda_{3a} + \lambda_6 = 0$ \\
Otherwise: $\lambda_{2a} - \lambda_{3a} + \lambda_6 = 0$.
\item From the stationarity condition with respect to \(\mathsf{r}_j\): $\lambda_4 - \lambda_5 = \frac{\nu \mathsf{r}^{\nu-1}_j}{\mathcal{E}_j} \mathbb{Q}_{ji}$. Given that \(\mathbb{Q}_{ji} = \max_{a \in V} (\mathsf{h}_{ja})\), we can determine that \(\mathbb{Q}_{ji}\) is maximized when \(\mathsf{h}_{ja}\) is at its upper bound \(\mathcal{R}_{ja}\): $\mathbb{Q}_{ji} = \mathcal{R}_{ja} \quad \text{where} \quad \mathsf{h}_{ja} = \max_{a \in V} (\mathsf{h}_{ja})$.

\end{itemize}
f) \emph{Summary of Optimal Values}:
\begin{itemize}
    \item $\mathbb{Q}_{ji} = \mathcal{R}_{ja}$
    \item $\mathsf{h}_{ja} = \mathcal{R}_{ja}$ if $\mathsf{h}_{ja} = \max_{a \in V} (\mathsf{h}_{ja})$
\end{itemize}
Thus, the optimal solution to the given problem is to set \(\mathbb{Q}_{ji}\) equal to the maximum allowable rate \(\mathcal{R}_{ja}\) for the flow \(\mathsf{h}_{ja}\) and ensure all constraints are satisfied.
\end{appendices}

\bibliographystyle{IEEEtran}

\end{document}